\documentclass[twoside]{article}
\usepackage{fleqn,espcrc2,amsmath}

\newcommand{\beeq}{\begin{equation}}
\newcommand{\eneq}{\end{equation}}
\newcommand{\beeqa}{\begin{eqnarray*}}
\newcommand{\eneqa}{\end{eqnarray*}}

\newcommand{\bma}{\begin{displaymath}}
\newcommand{\ema}{\end{displaymath}}

\usepackage{graphicx}

\newfont{\Bbb}{msbm10 scaled 1200}

\newfont{\Got}{eufm10 scaled 1200}
\def\permu{{\hbox{\Got S}}}

\def\calP{{\cal P}}
\def\calS{{\cal S}}

\def\calT{{\cal T}}

\def\FG#1{\textsf{FG}_{#1}}
\def\OFG#1{\FG{#1}^{+}}

\def\sgn{{\rm sgn}}

\def\idx#1{{\scriptscriptstyle (\! #1 \!)}}
\def\Balpha{{\mathbf{\alpha}}}

\title{Matrix model formulation of four dimensional gravity}

\author{Roberto De Pietri\address{Dipartimento di Fisica,
        Universit\`a di Parma and INFN gruppo collegato di Parma,
        via Parco Area delle Scienze 7/A, I-43100 PARMA, ITALY}
\\[0.5em]}

\begin{document}

\begin{abstract}

The attempt of extending to higher dimensions the matrix model
formulation of two-dimensional quantum gravity leads to the
consideration of higher rank tensor models. We discuss how these
models relate to four dimensional quantum gravity and the precise
conditions allowing to associate a four-dimensional simplicial
manifold to Feynman diagrams of a rank-four tensor model.
\vspace{1pc}
\end{abstract}

\maketitle
\section{INTRODUCTION}

The problem of constructing a quantum theory of gravity has been
tackled with very different strategies.  An attractive possibility is
that of encoding all possible space-times as specific Feynman diagrams
of a suitable field theory as it happens for the matrix model
formulation of two-dimensional quantum gravity (see for example
\cite{2d} and references therein). In the perturbative approach to the
matrix model the resulting Feynman diagrams have vertices which
correspond to two-simplices, and propagators which correspond to
edge-pairings, so a diagram leads to a surface obtained by glueing
triangles.  Indeed one is brought to the search for theories having
Feynman diagrams in which vertices can be identified with
$n$-simplices, and propagators with glueings of codimension-1
faces. If this happens, each Feynman diagram can be identified as
$n$-dimensional simplicial complex. We will discuss how the Feynman
diagrams of an $n$-tensor model can be interpreted in this
way. Moreover, we will discuss, in dimension four, the condition that
must be fulfilled in order that the resulting space is a four
manifold \cite{Conditions}.

\begin{figure*}
\centerline{\includegraphics[height=2.1cm]{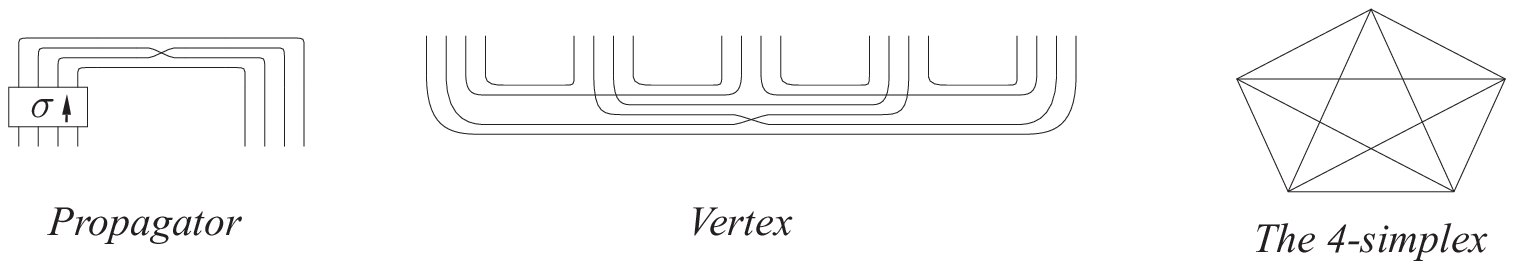}}
\vskip -1cm
\caption{\label{4tens} Feynman rules of a four-tensor generalized
matrix model and the analogy of the vertex diagram with the four
simplex.}
\end{figure*}

\section{GENERALIZED MATRIX MODELS}

An n-tensor model is a generalization of the matrix model where the 
basic configuration variable is an n-tensor fulfilling 
the symmetry condition
$\phi_{\alpha_{\tau(1)} \ldots \alpha_{\tau(n)} }
  =    \Re [ {\phi_{\alpha_{1}\ldots\alpha_{n}}} ]
   + i\cdot \sgn(\tau)\cdot \Im [{\phi_{\alpha_{1}\ldots\alpha_{n}}} ]$  
(where $\tau\in\permu_n$ and $\sgn(\tau)$ is the signature
(also called parity) of $\tau$) and partition function
\begin{gather}\label{Z:ntensor}
Z_n[N,\lambda] = \int [d\phi] \;\exp \Bigg[
      -\frac{1}{2} \sum_\Balpha |\phi_\Balpha|^2 
     \qquad
\\ \nonumber
 \quad +\frac{\lambda}{n+1}
        \!\sum_{\Balpha^\idx{0}\ldots \Balpha^\idx{n}}\!
         V^{\Balpha^\idx{0}\ldots \Balpha^\idx{n}}
         \cdot\phi_{\Balpha^\idx{0}}\cdot\ldots\cdot \phi_{\Balpha^\idx{n}}
       \Bigg]
\end{gather}
where $V^{\Balpha^\idx{0}\ldots \Balpha^\idx{n}}$ is a given
vertex function and multi-indices $\Balpha=(\alpha_1\ldots\alpha_n)$)
are used. Its Feynman diagram expansion is
\begin{align} \label{feyn:ntensor}
Z_n& [N,\lambda] = \sum_k 
                   \sum_{\sigma\in\permu_{\{0,\ldots,kn+k-1\}}}
\\ \nonumber 
   &    \frac{1}{k!} 
        \frac{\lambda^k}{(n+1)^k}
        \frac{(1/2)^{k(n+1)/2}}{(k(n+1)/2)!}
        \ \times 
\\ \nonumber 
   &  \times
        V^{\Balpha^\idx{0}\ldots \Balpha^\idx{n}}\cdot \ldots\cdot
        V^{\Balpha^\idx{kn+k-n-1}
           \ldots \Balpha^\idx{kn+k-1}}
       \ \times 
\\ \nonumber
   &  \times
        G_{\Balpha^\idx{\sigma(0)}\Balpha^\idx{\sigma(1)}}
        \cdot\ldots\cdot
        G_{\Balpha^\idx{\sigma(kn+k-2)}
           \Balpha^\idx{\sigma(kn+k-1)}}
\end{align}
and the propagator is given by:
\begin{equation} \label{eq:GMMprop}
G_{\alpha_1\ldots\alpha_n;\beta_1\ldots\beta_n}
   = \frac{2}{n!} 
        \!\!\!\!
        \sum_{\begin{array}{c} \scriptstyle \tau\in\permu_n \\[-1mm]
                               \scriptstyle \sgn(\tau)=-1 \end{array}}
        \!\!\!\!\!\!
        G_{\alpha_1\ldots\alpha_n;\beta_1\ldots\beta_n}^\idx{\tau},
\end{equation}
where $G_{\alpha_1\ldots\alpha_n;\beta_1\ldots\beta_n}^\idx{\tau}
= \delta_{\alpha_{\tau(1)}\beta_1} \ldots \delta_{\alpha_{\tau(n)}\beta_n}$.
It is straight-forward to encode the Feynman diagram expansion
(\ref{feyn:ntensor}) in terms of $(n+1)$-valent graphs as follows. 
We associate tensor expressions to vertices and edges of a graph, 
according to the following rules:
\begin{align}\label{graph:to:tensor:rules1}
   G^\idx{\tau}_{\Balpha^\idx{j}\Balpha^\idx{i}}
&\Rightarrow 
   \begin{array}{c}\setlength{\unitlength}{.75 pt}
   \begin{picture}(70,35)
   \put( 0, 0){\makebox(10,10){$\idx{j}$}}
   \put(20,10){\oval(30,30)[tl]}
   \put(20,25){\vector(1,0){30}}
   \put(30,27){\makebox(10,10){$\sigma$}}
   \put(50,10){\oval(30,30)[tr]}
   \put(60, 0){\makebox(10,10){$\idx{i}$}}
   \end{picture}\end{array}
\\ 
\label{graph:to:tensor:rules2}
   V^{\Balpha^\idx{i_0}\ldots \Balpha^\idx{i_n}}
&\Rightarrow  
   \begin{array}{c}\setlength{\unitlength}{.75 pt}
   \begin{picture}(80,35)
   \put(40,15){\oval(70,30)[b]}
   \put(40,15){\oval(40,30)[b]}
   \put(40, 0){\circle*{3}}
   \put(35,10){\makebox(10,10){$\ldots$}}
   \put( 5, 7){\makebox(10,10){$\scriptstyle 0$}}
   \put(20, 7){\makebox(10,10){$\scriptstyle 1$}}
   \put(75, 7){\makebox(10,10){$\scriptstyle n$}}
   \put( 0,20){\makebox(10,10){$\idx{i_0}$}}
   \put(15,20){\makebox(10,10){$\idx{i_1}$}}
   \put(70,20){\makebox(10,10){$\idx{i_n}$}}
   \end{picture}\end{array}
\end{align}
where $\tau=\sigma\circ (1\;n)$ and $\sigma$ is now an odd
permutation. We call the graphs $G\in\OFG{n}$ obtained using 
this procedure oriented {\it fat $n$-graphs}.


Using the definition just given we can rewrite (\ref{feyn:ntensor})
as a sum over all oriented fat $n$-graphs.  
Denoting by $\nu_0(G)$ and $\nu_1(G)$ the numbers
of vertices and edges of a fat graph $G$, we have
\begin{align} \label{Z:fatgraph}
Z_n[N,\lambda] &= 1+\!\! \sum_{G\in\OFG{n}} w_n(G)\cdot 
     \lambda^{\nu_0(G)} \cdot [[G]], \\
w_n(G) & =  \frac{\mu(G)}{
                   \nu_0(G)!\cdot (n+1)^{\nu_0(G)} \cdot (n!)^{\nu_1(G)}},
\nonumber
\end{align}
where $\mu(G)$ is the number of the inequivalent ways of labeling the 
vertices of $G$ with $\nu_0(G)$ symbols and $[[G]]$ (the weight factor)
denote the sum over all the possible values of the 
multi-indices $\Balpha^\idx{i}$ of the associated
tensor expression.

\section{ASSOCIATED COMPLEXES}

Consider a fat graph $G\in\OFG{n}$ and associate to 
each vertex of $G$ an $n$-simplex $S(v)$ with 
labeled vertices  $p_i(v)$ ($i=0,\ldots,n$) and the 
following object:
\begin{gather} \nonumber
   \!\!\!\!\!\!\!\!\!\!\!\!
   \begin{array}{c}\setlength{\unitlength}{.75 pt}
   \begin{picture}(250,50)
   \put(28,15){\makebox(0,0)[bl]{$\theta(p_0(v)$)}}
   \put( 5,30){\line( 1, 0){40}}
   \put( 5,35){\line( 0,-1){ 5}}
   \put( 0,37){\makebox(10,10){$\scriptstyle i_1^0$}}
   \put(15,35){\line( 0,-1){ 5}}
   \put(10,37){\makebox(10,10){$\scriptstyle i_2^0$}}
   \put(25,37){\makebox(10,10){$\ldots$}}
   \put(45,35){\line( 0,-1){ 5}}
   \put(40,37){\makebox(10,10){$\scriptstyle i_n^0$}}
   \put(98,15){\makebox(0,0)[bl]{$\theta(p_1(v)$)}}
   \put(75,30){\line( 1, 0){40}}
   \put(75,35){\line( 0,-1){ 5}}
   \put(70,37){\makebox(10,10){$\scriptstyle i_1^1$}}
   \put(85,35){\line( 0,-1){ 5}}
   \put(80,37){\makebox(10,10){$\scriptstyle i_2^1$}}
   \put(95,37){\makebox(10,10){$\ldots$}}
   \put(115,35){\line( 0,-1){ 5}}
   \put(110,37){\makebox(10,10){$\scriptstyle i_n^1$}}
   \put(228,15){\makebox(0,0)[bl]{$\theta(p_n(v)$)}}
   \put(205,30){\line( 1, 0){40}}
   \put(205,35){\line( 0,-1){ 5}}
   \put(200,37){\makebox(10,10){$\scriptstyle i_1^n$}}
   \put(215,35){\line( 0,-1){ 5}}
   \put(210,37){\makebox(10,10){$\scriptstyle i_2^n$}}
   \put(225,37){\makebox(10,10){$\ldots$}}
   \put(245,35){\line( 0,-1){ 5}}
   \put(240,37){\makebox(10,10){$\scriptstyle i_n^n$}}
   \put(125,0){\circle*{4}}
   \put(125,30){\oval(200,60)[b]}
   \put(125,30){\oval(60,60)[b]}
   \put(125,30){\oval(100,60)[br]}
   \put(160,20){$\cdots$}
   \end{picture}\end{array}
\\ \label{rules1}
\end{gather}
where $\theta(p_i(v))$ represents the face opposite to $p_i(v)$
and the sequence $(i_1^k,i_2^k, \cdots ,i_n^k)$ depends on whether
$n\cdot k$ is even or odd. If $n\cdot k$ is even then the sequence is
$(k-1,k-2,\cdots,k-n)$, with indices meant modulo $n+1$, while if
$n\cdot k$ is odd then the sequence is $(k+1,k+2,\cdots,k+n)$, with
indices again modulo $n+1$. 
Now, each edge of $G$ determines a pairing (simplicial identification)
between the $(n\!-\!1)$-faces associated to its ends. In fact
an edge of $G$ can be pictured as follows:
\begin{equation} \label{rules2}
   \begin{array}{c}\setlength{\unitlength}{.75 pt}
   \begin{picture}(200,60)
   \put(25, 5){\line( 0, 1){10}}
   \put(27, 0){$\theta(p_{j_0}\!(v))$}
   \put( 5,15){\line( 1, 0){40}}
   \put( 5,20){\line( 0,-1){ 5}}
   \put( 0,22){\makebox(10,10){$\scriptstyle j_1$}}
   \put(15,20){\line( 0,-1){ 5}}
   \put(10,22){\makebox(10,10){$\scriptstyle j_2$}}
   \put(25,22){\makebox(10,10){$\ldots$}}
   \put(45,20){\line( 0,-1){ 5}}
   \put(40,22){\makebox(10,10){$\scriptstyle j_n$}}
   \put( 40,35){\oval(30,30)[tl]}
   \put( 40,50){\vector(1,0){70}}
   \put( 70,52){\makebox(10,10){$\sigma$}}
   \put(110,35){\oval(30,30)[tr]}
   \put(125, 5){\line( 0, 1){10}}
   \put(127, 0){$\theta(p_{i_0}\!(w))$}
   \put(105,15){\line( 1, 0){40}}
   \put(105,20){\line( 0,-1){ 5}}
   \put(100,22){\makebox(10,10){$\scriptstyle i_1$}}
   \put(115,20){\line( 0,-1){ 5}}
   \put(110,22){\makebox(10,10){$\scriptstyle i_2$}}
   \put(125,22){\makebox(10,10){$\ldots$}}
   \put(145,20){\line( 0,-1){ 5}}
   \put(140,22){\makebox(10,10){$\scriptstyle i_n$}}
   \end{picture}\end{array}
\end{equation}
and it defines the map from $\theta(p_{i_0}(v))$ to $\theta(p_{j_0}(w))$ which
maps $p_{i_k}(v)$ to $p_{j_{\tau(k)}}(w)$ where $\tau= \sigma\circ
(1\;n)$. Summing up, we have associated to $G\in\OFG{n}$ a set $\calS$
of $n$-simplices and a face-pairing $\calP$ on this set. The result is
then a triangulated complex $X=\calS/\calP$ made up of glued
$n$-simplices. This is nothing else then the straightforward
generalization to arbitrary dimension of the rule used in the case of
the standard matrix model.  In fact, these rules associate 
the three basic order two diagram of the matrix
model:
\\[1mm]

\noindent
\mbox{\includegraphics[width=7cm]{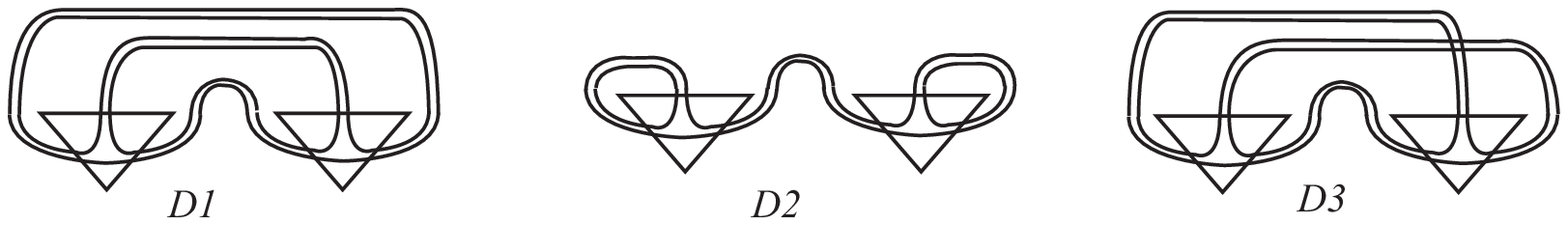}}

\noindent
to the torus (D3) and to the two inequivalent simplicial 
decomposition of the sphere (D1 and D2).

Taking Fig.~\ref{4tens} as model, one can rather easily 
transform (\ref{rules1}) and (\ref{rules2}) 
into  rules which allow to associate to a fat
graph $G$ a pattern of circuits on the graph. If there 
are $\nu_2(G)$ of these circuits and we
attach $\nu_2(G)$ discs $D^2$ to $G$ along them, we get a
two polyhedron that is the 2-skeleton of the cellularization dual
to the triangulation defined by the graph. We can then interpret the 
fat graph  as a way of describing the dual 2-skeleton of a 
triangulation. In particular, this dual 2-skeleton
determines the triangulation itself\footnote{In general, a
simplicial complex is {\em not} determined by the 2-skeleton dual
to the decomposition into simplices. This is however true if the
complex is obtained by glueing codimension-1 faces of simplices,
as in the case of complexes defined by fat graphs.}.
This consideration implies that a model was Feynman diagrams
can be coded in terms of fat graph can be seen as spin-foam 
models \cite{SpinFoam} and {\it viceversa}. Moreover,
if the vertex function, as in the case of the 4-tensor 
model defined by
\begin{align}
Z_4[N,&\lambda] = \!\!\int\! [d\phi]
   \exp\Bigg[\!-\frac{1}{2}\!\! \sum_{\alpha_1,\ldots,\alpha_4} \!\!
                   |\phi_{\alpha_1\alpha_2\alpha_3\alpha_4}|^2
\nonumber \\
  &+\frac{\lambda}{5}
         \sum_{\alpha_1,\ldots,\alpha_{10}}
         \phi_{\alpha_1\alpha_2\alpha_3\alpha_4}
         \phi_{\alpha_4\alpha_5\alpha_6\alpha_7}
         \phi_{\alpha_7\alpha_3\alpha_8\alpha_9}
\nonumber \\
& \quad\quad\quad\quad\quad
         \phi_{\alpha_9\alpha_6\alpha_2\alpha_{10}}
         \phi_{\alpha_{10}\alpha_8\alpha_5\alpha_1}
     \Bigg]
\ \ ,
\end{align}
is modeled on rule (\ref{rules1}), then, in the
evaluation of the weight factor $[[G]]$ of (\ref{Z:fatgraph}),
there are exactly $\nu_2(G)$ traces. Indeed
\begin{align}
Z_n&[N,\lambda] = 1+\!\!  \sum_{G\in\OFG{n}} 
            w_n(G)\cdot \lambda^{\nu_0(G)}\cdot N^{\nu_2(G)} 
\nonumber \\
   &= 1+\!\!  \sum_{G\in\OFG{n}} w_n(G) 
            {\rm e}^{-k_n \nu_n(\calT) + h_{n-2} \nu_{n-2}(\calT)}
\nonumber \end{align}
where, in the last line, we have introduced the standard dynamical 
triangulation constant 
and $\calT$ is the simplicial complex associated to the fat 
graph $G$.

\section{\label{cond} MANIFOLD CONDITIONS}

\begin{figure}
\centerline{\includegraphics[height=4cm]{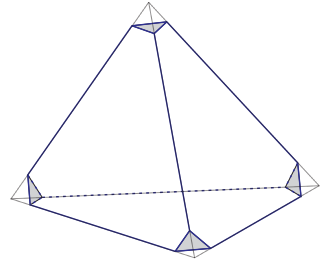}}
\vskip -1cm
\caption{\label{tetra} The basic building block in three-dimension.
The shaded areas, after gluing, will become the boundary components 
of $X^\partial$}
\end{figure}

In the previous section we discussed how to each fat graph is naturally
associated a simplicial complex obtained by orientation preserving 
gluing of simplices. We have associated to each $G\in\OFG{n}$ the
topological (triangulated) space X. There is indeed a very important
question to answer: is the space $X$ a manifold?  That is, it is true
that each point of $X$ has a closed neighborhood topologically equivalent 
to the $n$-disk $D^n$? In dimensions two, three and four (the only ones for 
which a definite answer it is available) the manifold question become 
simpler if we consider the closed space with boundary $X^\partial$ 
construct glueing the polyhedrons (instead of $n$-simplices) obtained 
removing the open star of the original vertices (as in fig.\ \ref{tetra}).
Then, $X$ is a manifold if and only if the boundary of $X^\partial$ is the
disjoint union of $(n\!-\!1)$-spheres. 

Clearly, there is nothing to check for all the points that lies on the
interior of the simplices or on codimension 1 faces. Indeed, in
dimension two, $X^\partial$ is always a manifold with boundary.
Moreover, since the boundary components are always circle, $X$ is 
always a Manifold. In dimension three, one has to check the manifold 
conditions only on the points lying on the edges. It comes out that, 
since we are considering only orientation preserving gluing, 
that $X^\partial$ is always a three manifold with boundary.

\begin{figure}
\centerline{\includegraphics[width=8cm]{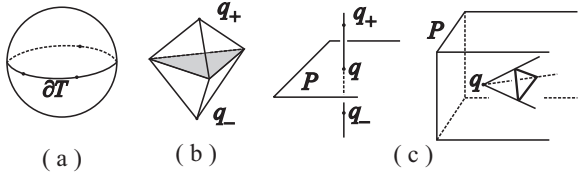}}
\vskip -1cm
\caption{\label{links} Links of the barycentres of a triangle (a) and 
an edge (b) in dimension four.  The link of the midpoint of an edge is 
the double cone on the link in a cross-section (c). The triangle involved 
in the {\it Surf condition} is shaded.}
\end{figure}
In dimension four the manifold question for $X^\partial$ has a more 
elaborate answer. In this case, we have to check the manifold condition 
on the barycenters of triangles and edges. They generate  
conditions {\em Cycl}\ and {\em Surf}\ of \cite{Conditions}, respectively.
It is important to note that they are purely combinatorial
conditions on the fat 4-graph.
By lack of space we can not give here a complete description
of these conditions and we refer the interested reader to 
\cite{Conditions}. They generate as follows. In PL-topology 
the concept of boundary of a closed neighborhood of a point $x$ is 
expressed as the link of the point $x$. We have that the manifold 
condition is indeed that the link of every points is homeomorphic 
to a 3-sphere. Now, we have that each 4-simplex contributes
to the link of a point on a triangle or on an edges 
with the components showed in Fig.\ \ref{links}. The gluing 
instruction translate on gluing instruction for these
components and the two conditions are the conditions that
the objects obtained after gluing are 3-spheres.

\section{GENERALIZED MODEL IN DIMENSION FOUR}

Since in dimension four, not to all the fat graphs is associated
a manifold, the $4$-tensor model cannot be used to define a viable theory 
of quantum gravity. We need a theory able to discriminate
fat graphs to which is associated a manifold with respect to the
other ones. This leads to consider generalization of this kind 
of model using fields over homogeneous (like Lie groups) spaces 
instead of $n$-tensors. The hope is that they are reach enough to make 
such discrimination. Example of theories of this kind are the ones 
discussed by Boulatov\footnote{The weight factor of 
this model \cite{Boulatov:1992} corresponds to the Ponzano-Regge 
\cite{Ponzano-Turaev-Viro} model and indeed much related to 
three dimensional euclidean gravity.} \cite{Boulatov:1992}, 
Ooguri \cite{Ooguri:1992b}, and more recently by De 
Pietri {\em et all.} \cite{4dimTM}. In dimension four one can use as 
basic variable a field $\phi(x_1,\ldots,x_4)$ over four copy of an homogeneous 
space $X$ over which is defined the action of 
a group $G$. It is possible to construct generalization of the 4-tensor 
model requiring that the field 
$\phi(x_1,\ldots,x_4)$ be real and invariant under any cyclic 
permutations of any three of its indices and using the action:
\begin{equation}
\begin{split}
&S[\phi] = \frac{1}{2} \int \prod_{i=1}^{4} dx_i ~ 
          \phi^{2}(x_1,x_2,x_3,x_4) 
\label{action} \\[-1mm] &
   + \frac{\lambda}{5!} \int \prod_{i=1}^{10} dx_i 
    ~~\phi(x_1,x_2,x_3,x_4) \phi(x_4,x_5,x_6,x_7)
\nonumber \\[-1mm]  &  
   \phi(x_7,x_3,x_8,x_9) \phi(x_9,x_6,x_2,x_{10}) 
   \phi(x_{10},x_8,x_5,x_1).
\nonumber
\end{split}
\end{equation}
In the case $X=G=SU(2)$ it is possible to show \cite{Ooguri:1992b}
that the Feynman diagram expansion of this theory
is still given by (\ref{Z:fatgraph}) where now 
the weight factor associated to each fat graph
$[[G]]=OCY[G]$ is the Ooguri-Crane-Yetter invariant 
construct on the dual two skeleton of the space
of glued simplices $G$. In the same way, 
if $X=SO(4)/SO(3)$, $G=SO(4)$ the same procedure will give 
a weight factor  $[[G]]=BC[G]$ is the Barrett-Crane \cite{Barrett:1998}
state sum associated to the space of glued simplices 
$G$ (see \cite{4dimTM}).



{\bf Acknowledgements:} This work is based on results obtained in
collaboration with Laurent Freidel, Kiril Krasnov, Carlo Petronio 
and Carlo Rovelli.



\end{document}